\documentclass[%
 reprint,
 amsmath,amssymb,
 aps,
prl,
]{revtex4-2}

\usepackage{graphicx}
\usepackage{dcolumn}
\usepackage{bm}

\usepackage{amsmath,amsfonts,amsbsy,amssymb,array,accents,dsfont}
\usepackage{enumerate,latexsym,graphicx}
\usepackage{xcolor,tikz}
\usepackage{tcolorbox}

\usepackage{subfigure}
\usepackage{stmaryrd}

\usepackage[colorlinks=true,      linkcolor=blue,      urlcolor=blue,      
            filecolor=blue,      citecolor=blue,       pdfstartview=FitH,     
						pdfpagemode=UseNone,      bookmarksopen=true]{hyperref}  
\usepackage[all]{hypcap}     

\newcommand{\beq}{\begin{equation}}
\newcommand{\eeq}{\end{equation}}
\newcommand{\bea}{\begin{eqnarray}}
\newcommand{\eea}{\end{eqnarray}}

\newcommand{\vep}{\varepsilon}

\newcommand{\der}{\partial}

\newcommand{\nn}{\nonumber}

\newcommand{\n}{{\mathsf n}}
\newcommand{\m}{{\mathsf m}}
\renewcommand{\k}{{\mathsf k}}

\usepackage{braket}
\usepackage{tikz}
\usetikzlibrary{matrix,calc,positioning,decorations.markings,decorations.pathmorphing,decorations.pathreplacing}
\usetikzlibrary{arrows,cd}
\usepackage{bbding}
\usetikzlibrary{positioning}
\tikzset{>=stealth}

\newcommand{\SSS}{\mathbb{S}}

\newcommand{\ZZ}{\mathbb{Z}}

\newcommand{\TT}{\mathbb{T}}
\newcommand{\Aa}{\mathcal{A}}

\newcommand{\Xx}{\mathcal{X}}

\newcommand{\Ww}{\mathcal{W}}

\newcommand{\R}{\ensuremath{\mathbb{R}}}

\newcommand{\xb}{\bar{x}}

\newcommand{\w}{\omega}

\usepackage{color}

\usepackage[normalem]{ulem}  

\definecolor{darkred}{rgb}{0.6,0,0}
\definecolor{darkblue}{rgb}{0,0,0.6}

\newcommand{\be}{\begin{equation}}
\newcommand{\ee}{\end{equation}}

\usepackage[bbgreekl]{mathbbol}
\usepackage{amsfonts}
\DeclareSymbolFontAlphabet{\mathbb}{AMSb} 
\DeclareSymbolFontAlphabet{\mathbbl}{bbold} 

\begin{document}

\preprint{APS/123-QED}

\title{
Worldsheet Correlators in Black Hole Microstates
}

\author{Davide Bufalini$^a$}
\author{Sergio Iguri$^{b,c}$}
\author{Nicolas Kovensky$^d$}
\author{David Turton$^a$}

\affiliation{ 
\vspace{0.2cm}
 $^{a}$Mathematical Sciences and STAG Research Centre, University of Southampton, Southampton
SO17 1BJ, United Kingdom,
\vspace{0.2cm}
}
\affiliation{
$^b$Instituto de Astronomía y Física del Espacio, CONICET-Universidad de Buenos Aires. C. C. 67, Suc. 28, 1428 Buenos Aires, Argentina,
\vspace{0.2cm}
}
\affiliation{
$^c$Mathematics with Computer Science Program, Guangdong Technion – Israel Institute of Technology, Shantou, Guangdong 515063, People's Republic of China,
\vspace{0.2cm}
}
\affiliation{
 $^d$Institut de Physique Th\'eorique, Universit\'e Paris Saclay, CEA, CNRS, Orme des Merisiers, 91191 Gif-sur-Yvette CEDEX, France.
 \vspace{0.2cm}
}

\date{August 18, 2022}

\begin{abstract}
Light probes interacting with heavy bound states such as black holes give rise to observables containing valuable dynamical information. 
Recently, a family of black hole microstates was shown to admit an exact string worldsheet description.
We construct the physical vertex operators of these models, and compute an extensive set of novel heavy-light correlators. 
We then obtain the first match between worldsheet correlators in black hole microstates and the holographically dual conformal field theory. 
We conjecture a closed formula for correlators with an arbitrary number of light insertions. 
As an application, we compute the analogue of the Hawking radiation rate for these microstates.
\end{abstract}

\maketitle

{\bf \textit{Introduction}}.---In recent years we have seen groundbreaking observations of black holes, as well as a sharpening of problems that arise in their quantum description, in particular the information paradox~\cite{Hawking:1976ra,Mathur:2009hf}. 
There are strong reasons to expect that black hole evaporation is unitary. However, a microscopic description of the quantum physics of general black holes, including singularity resolution and evaporation, is presently out of reach. 

In String Theory, black holes are bound states of strings and branes with an exponential number of internal microstates. 
Large classes of such microstates are well-described by smooth horizonless supergravity solutions~\cite{Lunin:2001jy,Bena:2016ypk,Bena:2017xbt,Heidmann:2019zws}, whose intricate horizon-scale structure appears black-hole-like for most purposes, while nevertheless allowing for potentially observable signatures~\cite{Bena:2020see,Bianchi:2020bxa}. Holography provides strong evidence for the microscopic interpretation of these solutions~\cite{Kanitscheider:2006zf,Giusto:2015dfa,Giusto:2019qig,Rawash:2021pik}.

Studying how light probes interact with a heavy background such as a black hole microstate yields valuable dynamical information, for instance about the unitary microscopic process underlying black hole evaporation. In holographic models, such processes can be understood in terms of correlators in the dual field theory involving two heavy operators and a number of light insertions, often referred to as heavy-light (HL) correlators~\cite{Galliani:2016cai,Fitzpatrick:2016mjq}.

While supergravity solutions are very useful descriptions, there is a wealth of string-theoretic physics that they do not capture. There are growing expectations that such stringy physics will be necessary to obtain a complete description of black hole microstructure. 
Recently, a novel set of worldsheet models describing black hole microstates has been constructed and studied~\cite{Martinec:2017ztd,Martinec:2018nco,Martinec:2019wzw,Martinec:2020gkv,Bufalini:2021ndn}. A subset of these models are exactly solvable gauged Wess-Zumino-Witten (WZW) models. They provide a rare and valuable window into stringy physics in highly-excited backgrounds such as black hole microstates.
In this Letter we construct the physical vertex operators of these models, and initiate the systematic study of their correlation functions. These describe the dynamics of perturbative strings in the given heavy backgrounds, and are exact in $\alpha'$. The main building blocks are given by correlators based on SL(2,$\R$) and SU(2). The former have a complicated structure due to spectral flow \cite{Maldacena:2000hw,Maldacena:2001km,Giribet:2007wp}, on which progress has been made recently \cite{Eberhardt:2019ywk,Dei:2021xgh}.

There are two instances of holography relevant to the models we study. 
On the gravity side, in the UV we have linear-dilaton NS5-brane asymptotics, and the dual theory is Little String Theory (LST)~\cite{Giveon:1999px,Giveon:1999tq}. Thus correlators of the vertex operators we construct holographically define LST amplitudes. In the IR, we have an AdS$_3$/CFT$_2$ duality. We will refer to the holographic CFT$_2$
as the HCFT.

Due to the gauging, the identification of the coordinates on the boundary of AdS$_3$ is subtle. By carefully making this identification, we derive all HLLH correlators 
where the light probes are massless,
in the AdS$_3$ limit. We find exact and highly non-trivial agreement with the small subset of known examples~\cite{Avery:2009tu,Galliani:2016cai,Lima:2021wrz}. We further identify the precise relation with the HCFT at the symmetric orbifold locus. Moreover, we conjecture a closed-form expression for all higher-point HL correlators on these backgrounds, Eq.\;\eqref{finalHLLLLLH} below.

As an application, we compute the emission rate for the unitary analogue of Hawking radiation from these backgrounds. Further applications of our results include studying the Penrose process \cite{Bianchi:2019lmi} and scrambling of infalling perturbations~\cite{Martinec:2020cml}. Our results also lay foundations for analyzing the stringy phenomenology of black hole microstates, which will be an important endeavour in the coming years.

\vspace{1.5mm}

{\bf \textit{Worldsheet models for black hole microstates}}.---We work in type IIB string theory on $\mathbb{R}^{4,1}\times \SSS^1 \times \TT^4$ with a microscopic $\TT^4$, and a macroscopic $\SSS^1$ with asymptotic radius $R_y$. We consider $n_5$ NS5-branes wrapped on $\TT^4 \times \SSS^1$, $n_1$ F1-strings wound on $\SSS^1$, and $n_p$ units of momentum charge along $\SSS^1$. We work in the NS5-brane decoupling limit, in which the string coupling $g_s$ and the radial coordinate $r$ scale to zero with $g_s/r$ fixed; we choose units in which $\alpha'=1$. 

We consider a set of horizonless supergravity solutions collectively known as spectral flowed supertubes. The non-BPS solutions in this family are known as the JMaRT solutions~\cite{Jejjala:2005yu}, and the family also contains BPS~\cite{Lunin:2004uu,Giusto:2004id,Giusto:2012yz} and/or two-charge solutions. These solutions can be recast in a simple form in terms of three integers $\m,\n,\k$, associated to the angular momenta and the orbifold structure near $r=0$, plus $R_y$~\cite{Bufalini:2021ndn}.

Flowing further to the IR, one can take the AdS$_3$ decoupling limit, after which the geometries are related by a large gauge transformation to global $(\mathrm{AdS}_3\times \SSS^3)/\mathbb{Z}_\k\times \TT^4$. The holographically dual CFT states are heavy in the sense that their conformal dimensions scale linearly with the central charge. 
At the locus in moduli space at which the HCFT is realized as the symmetric product orbifold Sym$^{N}(\TT^4)$ with $N=n_1n_5$, the states dual to the JMaRT solutions are related to the $\k$-twisted vacuum state by fractional spectral flow with parameters $(\m\pm\n)/\k$~\cite{Chakrabarty:2015foa}.

An exact worldsheet description of these configurations was constructed in \cite{Martinec:2018nco}. The target space is the coset
\begin{equation}
\label{eq: coset}
\frac{\mathrm{SL}(2,\R) \times \mathrm{SU}(2)  \times \R_t \times \mathrm{U}(1)_y}{\R \times \mathrm{U}(1)} \; \times \TT^4,
\end{equation}
where the gauging involves two null chiral currents, 
\begin{align}
\begin{aligned}
J &= J^3 + s_+ K^3 + i \mu \:\! \der t + i  k_+ \der y\,,\cr
\bar{J} &=  \bar{J}^3 + s_- \bar{K}^3 + i \mu \:\! \bar{\der} t + i k_- \bar{\der} y \,,\\[1mm]
s_\pm = & \, \n \pm \m \,, \quad\quad k_\pm=\mp \k R_y + n_5 \m \n /(\k R_y) \,,
\end{aligned}
\end{align}
where $J^3$ and $K^3$ are the Cartan generators of SL(2,$\R$) and SU(2), respectively, while $n_5 (1-s_\pm^2) +\mu^2 - k_\pm^2 = 0$ and $s_\pm \in 2\mathbb{Z}+1$. Supersymmetry is preserved in the left/right sector when $|s_\pm|=|k_\pm/\mu|=1$. The gauging effectively generates a $J\bar{J}$ marginal deformation on the worldsheet, similar to that of~\cite{Giveon:2017nie,Giveon:2017myj}.

\vspace{1.5mm}

{\bf\textit{Vertex operators}}.---The string spectrum contains both short and long strings, built upon the discrete and continuous series of SL(2,$\R$).
In this Letter, we shall primarily focus on 
supergravity operators, dual to light operators in chiral multiplets of the HCFT. The corresponding discrete series vertex operators with zero (worldsheet) spectral flows and $y$-winding are excitations of the center-of-mass wave-function $\Phi_{0} = V_{jm\bar{m}} V'_{j'm'\bar{m}'} e^{i(-E\,t + P_y y)}$, where $V_{jm\bar{m}}$ is a bosonic SL(2,$\R$) primary field of spin $j$ and $J^3$ eigenvalue $m$, and similarly for the SU(2) vertex $V'_{j'm'\bar{m}'}$ \cite{Kutasov:1998zh}, while $P_y = n_y/R_y$ with $n_y \in \mathbb{Z}$. First-excited states must satisfy the Virasoro and gauge constraints
\begin{align}
\label{eq:nullVirasoro}
0 \,=\, &\left[-j(j-1) + j'(j'+1)\right]/n_5 
    - \left(E^2 - P_{y}^2\right)/4\, ,  \\[1mm]  
0 &\,=\,  m + s_+ \, m' + \left(\mu E + k_+ P_{y}\right)/2 \, , \cr
0 &\,=\,  \bar{m} + s_-\, \bar{m}' + \left(\mu E + k_- P_{y}\right)/2 \, ,
\label{gaugecond}
\end{align}
which couple probe and background angular momenta. Sub-families of solutions to these conditions were analyzed in~\cite{Martinec:2017ztd,Martinec:2018nco,Martinec:2020gkv}.

We now construct the states polarized in the directions involved in the gauging. Importantly, the AdS$_3\times \SSS^3$ isometries are broken in the full model, and only emerge in the IR. Thus, vertex operators need not have definite SL(2,$\R$) and SU(2) spins or chiralities, which complicates the analysis. From now on we mostly write only holomorphic expressions, and assume normal ordering. The chiral BRST charge contains the usual terms plus additional contributions $\tilde{c} J$ and $\tilde{\gamma}\boldsymbol{\lambda}$ associated to the gauging;
here $\tilde{c}$, $\tilde{\gamma}$ are ghosts and the fermion $\boldsymbol{\lambda} = \psi^3 + s_+ \chi^3 + \mu \lambda^t + k_+ \lambda^y$ is the superpartner of $J$~\cite{Martinec:2020gkv}.

In the NS sector, before gauging (``upstairs'') we start from a generic linear combination of excitations in the  AdS$_3\times \SSS^3 \times \R_t \times \SSS^1_y$. Of these eight polarizations, two are removed by the $\gamma G$ and   $\tilde{\gamma}\boldsymbol{\lambda}$ constraints, $G$ being the supersymmetry current. Another two are BRST-exact combinations of $G \Phi_{0}$ and $\boldsymbol{\lambda} \Phi_{0}$. This gives the correct four transverse polarizations for a massless field in the six physical directions orthogonal to the gauging and the $\TT^4$. We obtain the physical operators 
\begin{align}
    \begin{aligned}
\Ww & = e^{-\varphi} \left(\psi_{\!_\perp} V_{j}\right)_{jm} V'_{j'm'} e^{ i (-E\,t +  P_y  y)}, \\
\Xx & = e^{-\varphi} V_{jm} \left(\chi_{\!_\perp} V'_{j'}
\right)_{j'm'} e^{i (-E\,t +  P_y  y)},
\end{aligned}
\end{align}
where $(\psi_{\!_\perp} V_{j})$ and $(\chi_{\!_\perp} V'_{j'})$ are generalizations of Eq.~(A.6) in \cite{Kutasov:1998zh} with the replacements $\psi^3 \to \psi^3 + c^t \lambda^t + c^y \lambda^y$ and $\chi^3 \to \chi^3 + d^t \lambda^t + d^y \lambda^y$, where
\begin{equation}
 c^t = \frac{-d^t}{s_+} = \frac{-n_5 P_y}{k_+ E + \mu P_y} \,, \quad~
 c^y = \frac{-d^y}{s_+} =   \frac{n_5 E}{k_+ E + \mu P_y} \,.
 \nonumber
\end{equation}
To take the AdS$_3$ limit, we define the rescaled energy ${\cal{E}} = E R_y$, and hold ${\cal{E}}$, $n_y$, and $s_{\pm}$ fixed as $R_y \to \infty$. The gauging is then concentrated mostly in $\R_t \times \mathrm{U}(1)_y$, and parametrically less in SL(2,$\R$)$\times$SU(2). As a result, $c^{t,y}, d^{t,y}$ become ${\cal{O}}\left(R_y^{-1}\right)$, and we recover the AdS$_3$ expressions.

In the R sector, before gauging, the polarizations involve a spin field that is a 64-component spinor of $O(10,2)$, $S_{\vep_1 \dots \vep_6}$ with $\vep_i=\pm 1$. Implementing GSO, we fix an overall spinor parity. The BRST procedure~\cite{Martinec:2020gkv}
reduces the d.o.f.~by a further factor of four, leaving the correct eight physical polarizations. Again, the $\TT^4$ does not participate in the gauging, so the corresponding signs $\vep_{4,5}$ are constrained only by overall parity.
Therefore, we make the ansatz
\begin{equation}
 \label{RRnullAnsatz}
	e^{-\frac{\varphi-\tilde{\varphi}}{2}}
	\sum_{\vep_1,\vep_2,\vep} 
	F_{\vep_1\vep_2}^{\vep} \,  
	S_{\vep_1 \dots \vep_6}  \, V_{j,m-\frac{\vep_1}{2}} \, V'_{j',m'-\frac{\vep_2}{2}} \, 
	e^{i(-E \:\! t + P_y \:\! y)} \,, \nn
\end{equation}
where $\vep_3 = \vep \vep_1 \vep_2$ and $\vep_6 = \vep \vep_4 \vep_5$, and solve for the coefficients $F_{\vep_1\vep_2}^{\vep}$. 
The field $\tilde{\varphi}$ arises from the bosonization of the $\tilde{\beta}\tilde{\gamma}$ ghosts for null gauging. In the $(-\frac{1}{2},\frac{1}{2})$ picture for $(\varphi,\tilde\varphi)$, there are no exact states. 
The BRST procedure mixes the different $\vep$-chirality sectors and leaves only two independent components in $F_{\vep_1\vep_2}^{\vep}$, say $F^+_{+\pm}$.
Acting with $\tilde{\gamma}\boldsymbol{\lambda}$ gives
\begin{align}\label{LambdaFrelation}
	\frac{F^-_{\vep_1 \vep_2 }}{F^+_{\vep_1 \vep_2 }} \,=\, 
	i \sqrt{n_5} \,  \frac{1 -  \vep_1 \vep_2 s_+}{ \mu + \vep_4  \vep_5  k_+}  \,=\, 
	\frac{i}{\sqrt{n_5}}  \, \frac{ \mu - \vep_4  \vep_5  k_+}{1 +  \vep_1 \vep_2 s_+}  \, , 
\end{align}
which allows us to decouple the remaining constraints.
The explicit relations will be given in a forthcoming companion work~\cite{Bufalini:2022}.
Finally, we can choose $F^\pm_{+\pm}$ by asking that the R states reduce to the usual AdS$_3$ ones in the IR. Interestingly, in this regime Eq.\;\eqref{LambdaFrelation} implies that $F^\pm  \sim 1$ and $F^\mp \sim {\cal{O}}\left(R_y^{-1}\right)$ for $\vep_4 \vep_5 = \pm 1$. We thus recover definite $\vep$-chirality states.

We derive the (worldsheet) spectral flowed versions of the above short string states analogously. For this, we use the same left/right spectral flow charge $\w$ in SL(2,$\R$) and SU(2). The Virasoro condition is modified accordingly, while the gauge conditions should be understood in terms of $m_w = \pm (J_\w + n)$, where $J_\w$ is the (flowed) spin and $n\in \mathbb{N}$. A residual discrete gauge symmetry allows us to recast $\w$ units of spectral flow in terms of $\k\w$ units of $y$-winding \cite{Bufalini:2021ndn}.

\vspace{1.5mm}

\textbf{\textit{Heavy-light correlators and spacetime basis.}}---
In the IR, correlators of the above vertex operators correspond to heavy-light $n$-point functions of the HCFT. In the remainder of this Letter we work in the AdS$_3$ limit.
The characteristics of the heavy states are encoded in the construction of the coset theory \eqref{eq: coset} itself. Hence, we can compute $\langle  O_H(0) O_L (1)  \bar{O}_L (x) \bar{O}_H (\infty) \rangle$ HCFT four-point functions by computing worldsheet two-point functions.

So far we have constructed vertex operators in the representation in which the Cartan currents are diagonalized, which provides the natural framework for our gauging procedure. 
Their correlators are the gauge-invariant subset of the so-called $m$-basis correlators of the upstairs theory, where the quantum numbers are related through Eq.\;\eqref{gaugecond}~\cite{Israel:2004ir,Chung:1992mj}. In particular the two-point functions factorize, the only non-trivial contributions being the SL(2,$\R$) propagators \cite{Maldacena:2001km}.

Holographic applications of AdS$_3$ string models involve the conjugate $x$-basis, in which the SL(2,$\R$) currents act as differential operators associated to the spacetime Virasoro generators. The complex label $x$ is identified with the coordinate of the HCFT.
More explicitly, in global AdS$_3$, given an operator $\mathcal{V}_{hm\bar{m}}$ of spacetime weight $h$, with $m=h+n$, $\bar{m}=h+\bar{n}$, $n,\bar{n} \in \mathbb{N}$, one defines
$\mathcal{V}_{h}(x) = \sum_{m,\bar{m}} x^{m-h}\bar{x}^{\bar{m}-h}\mathcal{V}_{hm\bar{m}}$~\cite{Maldacena:2000hw,Maldacena:2001km}. Here $\mathcal{V}_{hm\bar{m}}$ contains any excitations, such as fermions or spin fields. However, in the gauged models \eqref{eq: coset}, the identification of the $x$ variable is obscured by the gauging, since $J^\pm$ do not commute with the BRST charge.

We now show that a careful reconstruction of the local basis in these cosets leads to a considerable set of interesting new results, a subset of which match very non-trivially to two families of known HLLH correlators.

We choose a gauge in which the upstairs SL(2,$\R$) time and angular direction are fixed. Then, importantly, $t/R_y$ and $y/R_y$ parametrize the asymptotic boundary of the downstairs AdS$_3$ at a fixed point on the $\SSS^3$. We define $m_y = \left({\cal{E}} + n_y \right)/2$ and $\bar{m}_y = \left({\cal{E}} - n_y \right)/2$, and interpret these as the asymptotic AdS$_3$ mode labels of the cosets.

At large $R_y$, the modified Virasoro condition \eqref{eq:nullVirasoro} is solved by the usual relation $j=j'+1 + {\cal{O}}(R_y^{-2})$. Although the weights of the $t,y$-exponentials scale as ${\cal{O}}(R_y^{-2})$, the associated quantum numbers ${\cal{E}}$ and $n_y$ enter non-trivially in the leading-order gauge conditions \eqref{gaugecond} because $\mu \sim k_+ \sim -k_- \sim -\k R_y$.   Hence, the vertex operators reduce to the familiar AdS$_3$ ones, however their quantum numbers must satisfy the gauge conditions
\begin{equation}
\label{eq:gauge-my}
    0 \;=\; m + s_+ m' - \k \, m_y 
    \;=\; \bar{m} + s_- \bar{m}' - \k \, \bar{m}_y  \, .
\end{equation}
Therefore, $m_y$ and $\bar{m}_y$ take values in  $\mathbb{Z}/\k$. Crucially, momentum quantization along the $y$-circle imposes that $m_y-\bar{m}_y \in \mathbb{Z}$. This implies that local operators are built only out of the subset of SL(2,$\R$) modes that satisfy 
\begin{equation}
\label{eq:modk}
    m-\bar{m} \;\equiv\; s_+ m' - s_- \bar{m}' ~~~ (\mathrm{mod} \,\, \k).
\end{equation}
Since $m_y$ and $\bar{m}_y$ label the spacetime modes, we define the $x$-basis operators in the coset as 
\begin{align}
\begin{aligned}
\label{xbasisops}
    O_h(x) & = \frac{1}{\k^{2h}}  \sum_{m_y,\bar{m}_y}
    x^{m_y-h}\bar{x}^{\bar{m}_y-h} \mathcal{V}_{hm\bar{m}}\mathcal{V}_{h' m'\bar{m}'}' \,,
\end{aligned}
\end{align}
where $h'$ is related to $h$, and the left- and right-handed quantum numbers are constrained by Eqs.\;\eqref{eq:gauge-my}--\eqref{eq:modk}.
Coset HLLH correlators are accordingly expanded as power series in $x,\bar{x}$, where the coefficients are the $m$-basis two-point functions. 

We now have everything in place to compute a large set of correlators. We first present a simple example involving the untwisted $h=1/2$ chiral primary of the orbifold CFT, denoted by $O_L=O^{++}$. This is associated to a RR worldsheet state (which we also denote by $O^{++}$) for which the SL(2,$\R$) mode propagators are trivial. We write $s_+=2s+1$, $s_-=2\bar{s}+1$, which makes explicit that $s_\pm$ are odd, and connects to the notation of previous works. We decompose $s=p\k+\hat{s}$ and $\bar{s}=\bar{p}\k+\tilde{s}$ with $0\le \{\hat{s},\tilde{s} \} < \k$, and define $\delta = \tilde{s} - \hat{s}$. Then we obtain
\begin{align}
&\langle O^{++}(1)O^{--}(x)\rangle_H =
\label{eq:HLLHOpp}\\ 
&\qquad \frac{1}{\k^{2}}\frac{(1-x)|x|^{\delta/\k}
    + (1-\bar{x})|x|^{-\delta/\k}
    -|1-x|^2 |x|^{-|\delta|/\k}
    }{x^{(2s+\delta)/2\k} \, \xb^{(2\bar{s}-\delta)/2\k} \, |x||1-x|^2
    \left(1-|x|^{2/\k}\right)} 
    \,, \nonumber
\end{align}
where $\langle \dots \rangle_H$ stands for the correlator 
in the coset corresponding to the heavy backgrounds of interest. For the appropriate subset of values of $s_\pm$ and $\k$,
this worldsheet correlator matches the supergravity and orbifold CFT results of \cite{Galliani:2016cai,Avery:2009tu,Avery:2009xr}. It also significantly extends these results to the full set of non-BPS JMaRT backgrounds.

\vspace{1.5mm}

\textbf{\textit{HL correlators and the Sym$^{N}(\TT^4)$ CFT.}}---
By enforcing the constraint \eqref{eq:modk} using a Kronecker comb, $ \sum_{q \in \mathbb{Z}} \delta_{m-\bar{m},\k q} = \k^{-1} \sum_{r = 0}^{\k-1} e^{2 \pi i r \, \frac{m - \bar{m}}{\k}}$, the sum in \eqref{xbasisops} over \textit{restricted} powers of the physical cross-ratio $x$ becomes a sum over \textit{unrestricted} powers of its $\k^{\mathrm{th}}$ roots, i.e.~a sum over powers of all $u$ such that $u^\k=x$.
Then the RHS of Eq.\;\eqref{eq:HLLHOpp} takes the simple and elegant form
\begin{equation}
\langle O^{++}(1)O^{--}(x)\rangle_H \;\! =\;\!  \frac{1}{\k^{3}} \sum_{u^\k=x} \frac{u^{-\frac{s_+}{2}}\bar{u}^{- \frac{s_-}{2}}|u|^{1-\k}}{|1-u|^2} \,. 
\end{equation}
This formula is extremely instructive.
First, it is readily generalized to light insertions with generic weights and R-charges. Second, the appearance of the $\k^{\mathrm{th}}$ roots of $x$ invites a comparison with the dual HCFT at the symmetric orbifold point. We now develop both these points.

Let us consider a worldsheet operator with definite charge $m'$, and express the coset spacetime modes as $m_y = x\der_x + h$. Proceeding similarly for the \textit{upstairs} SL(2,$\R$) modes $m=u \der_u + h - \beta$, for the auxiliary variable $u$ and shift $\beta$, we see that in order to obtain $u^\k=x$, we should fix $\beta$ and the analogous right-moving $\bar\beta$ to be
\begin{align}\label{eq: beta}
   \beta  = h (1-\k) + s_+ m'  \, , \quad \bar{\beta} = h(1-\k) + s_- \bar{m}' \,.
\end{align}
Then the worldsheet operator \eqref{xbasisops} is re-expressed as  
\begin{align}
\begin{aligned}
\label{xbasisops-2}
    O_h(x) & \;\! = \;\! 
    \frac{1}{\k^{2h+1}}\sum_{u^{\k}\,=\,x} u^{\beta}\bar{u}^{\bar{\beta}}  \mathcal{V}_{h}(u) \mathcal{V}'_{h' m'\bar{m}'}  \,.
\end{aligned}
\end{align}
The role of $\beta$ is two-fold: it gives the Jacobian factor for the coordinate change from $x$ to $u$, and also the appropriate rescaling under spectral flow \cite{Avery:2009tu}.
We thus find that worldsheet HLLH correlators with generic massless insertions take the surprisingly simple form
\begin{equation}
\label{finalHLLHbeta}
    \langle O_L(x_1)\bar{O}_L(x_2) \rangle_{H}  \;\!=\;\! 
    \frac{1}{\k^{4h+2}} 
    \sum_{u_i^\k=x_i}
    \dfrac{u_1^{ \beta_1}\bar{u}_1^{ \bar{\beta}_1} u_2^{ \beta_2}\bar{u}_2^{ \bar{\beta}_2}   }{|u_{1}-u_{2}|^{4h}}  \, .
\end{equation}

\textit{A priori}, this gives a prediction for the HCFT at strong coupling. However, a subset of these correlators is known to be protected between supergravity and symmetric orbifold CFT~\cite{Galliani:2016cai,Avery:2009tu,Avery:2009xr}. Thus it is natural to investigate how general this protection may be.
For untwisted light operators of the orbifold CFT, it is natural to identify $u$ with the coordinate on the $\k$-fold covering space that trivializes the twist operators involved in the definition of the heavy states. This coincides precisely with the symmetric orbifold covering space method for such correlators~\cite{Lunin:2000yv,Lunin:2001pw}. In the presence of twist-$\k$ operators, modes take values in  $\mathbb{Z}/\k$. Moreover, when mapping $S_{\k}$-invariant untwisted operators to the $\k$-fold covering space, one obtains an expression analogous to Eq.\;\eqref{xbasisops-2}.

By contrast, for twisted operators, the interpretation of our result \eqref{finalHLLHbeta} is more involved: the covering map for such correlators is \textit{not} $x=u^\k$.
To exhibit the precise relation, we focus on the singlet marginal deformation operator of twist two, and take $s_\pm=0$. More general light operators and values of $s_\pm$ can be treated analogously.
The relevant four-point function was studied recently in 
\cite{Lima:2020nnx,Lima:2021wrz,Lima:2022cnq} 
in the Sym$^{N}(\TT^4)$ CFT. At large $N$ the correlator is dominated by a contribution from a covering space with genus zero.
The relation between the physical-space cross-ratio and its covering-space images is $x = (u+1)^{2\k}(u-1)^{-2\k}$. The relevant correlator takes the form $G(u) = u^{-2}(u+1)^{2+2\k}(u-1)^{2-2\k}$. The final result is obtained by summing over the $2\k$ pre-images of $x$, and including the appropriate combinatorial factor. However, due to the $u \to 1/u$ symmetry of the map, there are actually only $\k$ distinct contributions. Upon inserting the explicit solutions $u=u(x)$, the final expression remarkably coincides precisely with \eqref{finalHLLHbeta}.

From the point of view of the worldsheet theory, there is no strong distinction between vertex operators associated to different twist sectors of the HCFT at the orbifold point, and our general result depends only on their weight and charges. In this sense, the only \textit{covering space} that appears is that associated to the heavy operators $x = u^\k$.

Moreover, using Eq.~\eqref{xbasisops-2}, we conjecture a formula for worldsheet correlators with $n$ massless insertions. 
Denoting the light operator weights by $h_i$ and charges by $m_i', \bar{m}_i'$, and
defining the shorthands $O_i \equiv O_{h_i}$ and $\hat{O}_i(u_i) \equiv {\cal{V}}_{h_i}(u_i){\cal{V}}'_{h_i'm_i'\bar{m}_i'}$,
we obtain
\begin{align}
\label{finalHLLLLLH}
    \begin{aligned}
    & \langle O_{1}(x_1) \dots O_{n}(x_n) \rangle_H = \\
    & \frac{1}{\k^{2\mathsf{H} +n}} 
    \sum_{u_i^\k=x_i} 
    \bigg(\prod_{i=1}^n u_i^{\beta_i}
    \bar{u}_i^{\bar{\beta}_i} 
    \bigg)
    \langle \hat{O}_1(u_1) \dots \hat{O}_n(u_n)\rangle
   \,  , 
    \end{aligned}
\end{align}
with $\beta_i$, $\bar{\beta}_i$ as in Eq.\;\eqref{eq: beta}, and $\mathsf{H} =h_1 + \dots + h_n$. In this way, these HL correlators are directly determined from the \textit{vacuum} $n$-point function evaluated at the $\k^{\mathrm{th}}$ roots of the physical insertion points. 
Interestingly, this suggests that the large $N$ fusion rules for these correlators within our heavy states coincide with the vacuum ones, including those involving spectrally flowed states.

\vspace{1.5mm}

\textbf{\textit{Hawking radiation from the worldsheet}.}---
The asymptotically flat JMaRT solutions emit ergoregion radiation~\cite{Cardoso:2005gj}. This has been interpreted as an enhanced analogue of Hawking radiation, since both are described by the same microscopic process in the HCFT~\cite{Chowdhury:2007jx,Avery:2009tu}. As an application, we now rederive the amplitude controlling the rate of emission of supergravity quanta from the JMaRT solutions. This is obtained by taking $x_2\to 0$ and $x_1=x$ in Eq.~\eqref{finalHLLHbeta}, obtaining
\begin{equation}
\label{HawkingFinal}
	\Aa(x) \;\!=\;\!  \frac{1}{\k^{2h}} 
   \dfrac{\sum_{\ell \in \ZZ}\delta_{s_+ m' - s_-\bar{m}',\,  \k \; \! \ell} }{x^{h(1+\frac{1}{\k}) - m'\frac{s_+}{\k}} \; \bar{x}^{h(1+\frac{1}{\k}) - \bar{m}'\frac{s_-}{\k}}}\, \, .
\end{equation}
The numerator ensures the correct spectrum of emission. 
This amplitude gives a direct worldsheet derivation of the emission rates computed in  \cite{Chowdhury:2007jx,Avery:2009tu,Avery:2009xr,Chakrabarty:2015foa}.

\vspace{1.5mm}

\textbf{\textit{Discussion}.}---
We have obtained closed-form expressions for a comprehensive family of correlators of light operators in heavy black hole microstates, using the exact worldsheet description of the JMaRT solutions. We did so by carefully identifying the correct spacetime modes and the local coordinate on the asymptotic AdS$_3$ boundary emerging in the IR. 
For all cases computed in the literature,
our worldsheet results precisely match with those of the dual HCFT at the symmetric orbifold locus.

Our findings are a significant improvement over previous work, providing a powerful general method that is much less computationally intensive than the Lunin-Mathur covering space method in the symmetric orbifold CFT, as well as a set of results that substantially expand upon the small number of earlier case-by-case studies.

The matching to the HCFT for particular sub-cases of our results \eqref{eq:HLLHOpp} and \eqref{finalHLLHbeta} is highly non-trivial. 
Perhaps most strikingly, even for the non-supersymmetric amplitudes \eqref{HawkingFinal}, we find precise agreement. There is no known non-renormalization theorem protecting the latter amplitudes. However the matching we find is consistent with previous results, and is almost certainly due to the relation between these heavy states and the NS vacuum involving orbifolding and fractional spectral flow \cite{Chakrabarty:2015foa}.

Our results open up many interesting directions for future work. Eqs.~\eqref{finalHLLHbeta} and \eqref{finalHLLLLLH} can be used to obtain a deeper understanding of factorization channels, conformal blocks and anomalous dimensions of the HCFT~\cite{Lima:2021wrz,Lima:2022cnq}. 
Furthermore, the vertex operators we have constructed will enable the explicit study of  correlators in the full coset models. In the UV, the holographically dual LST is non-local, 
and we see that our construction of the $x$ coordinate breaks down, as it must. Computing these correlation functions might very well take us one step closer to flat space holography~\cite{Asrat:2017tzd}. 

Finally, our methods can be used to study a wealth of black hole microstate physics including the Penrose process \cite{Bianchi:2019lmi}, tidal forces \cite{Martinec:2020cml} and multipole ratios \cite{Bena:2020see}, all of which can be probed with our higher-point functions. 
Our work lays the foundations for computing more general heavy-light correlators from the string worldsheet, which will contribute to the emerging programme of exploring the stringy phenomenology of black hole microstates in this exciting era of direct detections of gravitational waves \cite{LIGOScientific:2016aoc}.

\section{Acknowledgments}
\begin{acknowledgments}
For discussions, we thank I.~Bena, S.~Chakraborty, N.~\v{C}eplak, A.~Dei, S.~Giusto, S.~Hampton, M.~Gra\~na, M.~Guica, Y.~Li, E.~Martinec, S.~Massai, S.~Rawash, R.~Russo and  M.~Santagata.
D.B.~is supported by Royal Society Research Grant RGF\textbackslash R1\textbackslash181019. N.K.~is supported by ERC Consolidator Grant 772408-Stringlandscape. 
D.T.~is supported by Royal Society Tata URF UF160203.
For hospitality, D.B thanks CEA Saclay and N.K. thanks the University of Southampton.
\end{acknowledgments}

\end{document}